\documentclass[12pt,preprint]{aastex}
\usepackage{color}
\newcommand{\be}{\begin{equation}}
\newcommand{\ee}{\end{equation}}

\newcommand{\msun}{{M}_{\sun}}

\slugcomment{Printed at \today}
\shorttitle{The universal ``heartbeat" oscillations in BHs} \shortauthors{Wu et al.}

\begin{document}

\title{The universal ``heartbeat" oscillations in black hole systems accross the mass-scale}

\author{Qingwen Wu\altaffilmark{1}, Bozena Czerny\altaffilmark{2,3}, Mikolaj Grzedzielski\altaffilmark{3}, Agnieszka Janiuk\altaffilmark{3}, Wei-Min Gu\altaffilmark{4}, Ai-jun Dong\altaffilmark{1}, Xiao-Feng Cao\altaffilmark{5}, Bei You\altaffilmark{2}, Zhen Yan\altaffilmark{6} and Mou-Yuan Sun\altaffilmark{4} }
\altaffiltext{1}{School of Physics, Huazhong University of Science and Technology, Wuhan 430074, China}
\altaffiltext{2}{Nicolaus Copernicus Astronomical Center PAS, Bartycka 18, 00-716 Warsaw, Poland}
\altaffiltext{3}{Center for Theoretical Physics, Al. Lotnikow 32/46, 02-668 Warsaw, Poland}
\altaffiltext{4}{Department of Astronomy, Xiamen University, Xiamen, Fujian 361005, China}
\altaffiltext{5}{School of Physics and Electronics Information, Hubei University of Education, 430205 Wuhan, China}
\altaffiltext{6}{Key Laboratory for Research in Galaxies and Cosmology, Shanghai Astronomical Observatory, Chinese Academy of Sciences, 80 Nandan Road, Shanghai 200030, China}

\begin{abstract}
The hyperluminous X-ray source (HLX-1, the peak X-ray luminosity $\sim 10^{42}\rm erg\ s^{-1}$) near the spiral galaxy ESO 243-49 is possibly the best  candidate for intermediate mass black hole (IMBH), which underwent recurrent outbursts with a period of $\sim 400$ days. The physical reason for this quasi-periodic variability is still unclear. We explore the possibility of radiation-pressure instability in accretion disk by modeling the light curve of HLX-1, and find that it can roughly reproduce the duration, period and amplitude of the recurrent outbursts HLX-1 with an IMBH of $\sim 10^5\msun$. Our result provides a possible mechanism to explain the recurrent outbursts in HLX-1. We further find a universal correlation between the outburst duration and the bolometric luminosity for the BH sources with a very broad mass range (e.g., X-ray binaries, XRBs, HLX-1 and active galactic nuclei, AGNs), which is roughly consistent with the prediction of radiation-pressure instability of the accretion disk. These results imply that ``heartbeat" oscillations triggered by radiation-pressure instability may appears in different-scale BH systems.    
\end{abstract}


\section{Introduction}
The central engines of black-hole (BH) systems are generally believed to be scale-free \citep[e.g.,][]{merloni03,falcke04,mchardy06,dong14}, with the stellar-mass-BH X-ray binaries (XRBs, 3-20$\msun$) being a scale-down version of active galactic nuclei (AGNs) containing supermassive BHs ($10^{6-10}\msun$). The evidence for intermediate mass BHs (IMBHs, $10^{2-5}\msun$) is still weak, although recently a hyperluminous X-ray source (HLX-1, the peak X-ray luminosity $\sim 10^{42}\rm erg\ s^{-1}$) near the spiral galaxy ESO 243-49 was found \citep[e.g.,][]{farrell09}, which may be powered by an IMBH with the mass $\sim 10^{4-5}\msun$ \citep[e.g.,][]{davis11,serv11,webb12,straub14}. 

The inner region of the geometrically thin, optically thick accretion disk (Shakura-Sunyaev disk, SSD, \citealt{shakura73}) will be dominated by the radiation rather than the gas pressure if the accretion rate is more than a few percent of the Eddington limit. Shortly after the $\alpha$-viscosity disk model was proposed, it was realized that radiation-dominated accretion flows are thermally unstable if the viscous torque is proportional to the total (gas plus radiation) pressure. In numerical experiments, \citet{hirose09} reported that the disk is thermally stable with flux-limited diffusion method in radiation-pressure dominated region. However, \citet{jiang13} found that this thermal instability does exist after using more advanced numerical methods (see \citealt{mishra16} for a similar conclusion). Observationally, GRS 1915+105 has shown spectacular X-ray variabilities, and among the different variability classes (which were referred to different Greek letters) the $\rho$ class shows regular, repetitive `heartbeat'-like oscillations in the light curve with recurrence time of $\sim 50$ s \citep[e.g.,][]{belloni00,fender04}. Recently, the similar variability patterns have also been found in IGR J17091-3624 \citep[e.g.,][]{alta11}. Such heartbeat oscillations can be nicely explained by the radiation-pressure instability that predicts the limit-cycle variations \citep[e.g.,][]{janiuk02}. \citet{bz15} found that a neutron star-MXB 1730-335 also exhibits large-amplitude quasi-periodic oscillations that are very similar to the ’heartbeat’ variability in GRS 1915+105 and IGR J17091-3624. Further more, it was suggested that the short-lived compact radio sources may also be triggered by the radiation-pressure instability since the ages of these compact radio sources are roughly consistent with the predictions of radiation-pressure instability model \citep[][]{czerny09,wu09}. The narrrow-line Seyfert I of IC 3599 that showed two high-amplitude outbursts in the last several tens year, which may also have its outbursts triggered by the radiation-pressure instability of the disk \citep[e.g.,][]{gru15}.

Monitoring observations of HLX-1 with \emph{Swift} X-ray satellite showed that the source underwent recurrent outbursts with a period of $\sim400$ days \citep[e.g.,][]{yan15}. The physical reason for this quasi-periodic variability is still unclear, considered possibilities include precession of a beamed X-ray jet \citep[][]{king14}, or a periodic enhancement of the accretion rate triggered by the passage at periastron of a donor star on an eccentric orbit \citep[e.g.,][]{lasota11,helm16}. \citet{lasota11} also investigate the possibilities of disk instabilities, where they rule out the ionization instability, but they do not discuss in detail the radiation pressure instability as a plausible mechanism. In this work, we explore the possible mechanisms of radiation-pressure instability for HLX-1 based on numerical calculations. We will further investigate the possibility of the universal properties of BH sources that possibly triggered by the radiation-pressure instability. In calculation of the luminosities in HLX-1 and AGNs , we assume the following cosmology: $H_{0}=70\ \rm km\ s^{-1} Mpc^{-1}$, $\Omega_{0}=0.3$ and $\Omega_{\Lambda}=0.7$.  
 
\section{Results and Discussions}
\subsection{Radiation-pressure instability and recurrent outbursts of HLX-1}
 
 We model the light curve of HLX-1 based on the time evolution of the standard thin disk under the radiation-pressure instability. For a general description of the viscous stresses $T_{r\phi}=-\alpha P_{\rm tot}^{\mu}P_{\rm gas}^{1-\mu}$, the standard thin disk is unstable at high mass accretion rate at small radii if $3/7<\mu\leq1$\citep[e.g.,][]{szus90}, where total pressure is gas and radiation pressure ($P_{\rm tot}= P_{\rm gas}+P_{\rm rad}$). For simplicity, we only consider the evolution of the disk in the present work, where the possible corona and jet/wind are neglected in order to keep the picture as simple as possible. The model is parameterized by BH mass, $M_{\rm BH}$, accretion rate, $\dot{M}$, viscosity parameter, $\alpha$, and a constant, $\mu$, where the reader is referred to \citet{czerny09} and \citet{janiuk11} for more details on the model. In this work, we adopt $\alpha=0.02$ as suggested from the previous studies of AGN variability \citep[][]{siem89,star04}.

 In Figure 1, we present the light curve of HLX-1. The \emph{Swift} X-ray telescope (XRT) monitored HLX-1 regularly from 2008 October 24, and we obtain the observational counts evolution in 0.3-10 keV band from the online \emph{Swift}/XRT product generator \citep[][]{evans09} by using a dynamical count binning of 20 counts per bin (see top panel of Figure 1).  In the bottom panel of Figure 1, we present the theoretical flux evolution of HLX-1, where $\mu=0.57$, $M_{\rm BH}=1.2\times10^{5}\msun$, $\dot{M}=5.7\times10^{-4}\msun/{\rm yr}$ (0.22 Eddington accretion rate) are adopted to reproduce the observational properties of the light curve for HLX-1. In the model, the heavier BH leads to a stronger variations (larger ratio of the maximum and minimum luminosities, $L_{\rm max}/L_{\rm min}$), and a longer outburst period ($P$). The BH mass of $M_{\rm BH}=1.2\times10^{5}\msun$ of HLX-1 is mainly determined by the ratio of $L_{\rm max}/L_{\rm min}\sim 30$ and the period of $P\sim400$ days from one outburst to another one. The average bolometric luminosity, $L_{\rm bol}$, of the model is controlled by the input accretion rate $\dot{M}$. We find that the parameter $\mu$ will affect the ratio of the outburst duration and the period, where $\mu\sim0.57$ is determined mainly from the ratio of $T_{\rm dur}/P\sim0.1$ for HLX-1, where the outburst duration $T_{\rm dur}$ is measured using the time between the luminosity increases and decreases to $(L_{\rm max}+L_{\rm min})/2$ respectively. Therefore, parameters of $M_{\rm BH}$, $\dot{M}$ and $\mu$ can be roughly constrained from the observational quantities of $L_{\rm bol}$, $P/T_{\rm dur}$ and $L_{\rm max}/L_{\rm min}$. Our results suggest that the recurrent outbursts in HLX-1 can be roughly reproduced by radiation-pressure instability in the SSD surrounding an IMBH.

A $\sim10^5\msun$ IMBH is needed in our outburst modeling, which is roughly consistent with that derived from spectrum modeling \citep[e.g.,][]{davis11,serv11,straub14}. We note that the simplest SSD model is adopted in modeling the light curve of HLX-1, and the BH mass estimation may differ a little if considering the possible disk corona and wind/jet, but our main conclusion will not change. The radiation-pressure instability model normally predicts the light curve with a slow rise and fast decay (SRFD) as observed in GRS 1915+105 and IGR J17091-3624. However, the light curve of HLX-1 do not show very evident SRFD profiles except the outburst during MJD=57000-57150 (see Figure 1). We note that this may be caused by the low-quality of X-ray data at dips of outbursts due to the sensitivity of \textit{Swift}, and the high-sensitivity observations will help to further explore this issue. The outbursts in HLX-1 are not strictly periodic (e.g., one-month delay for the outburst in 2013), which may be caused by the change of the accretion rate in each outburst due to the possible jet/wind and/or corona. Actually, GRS 1915+105 and IGR J17091-3624 show several different variability classes, and the period vary by several percent from one outburst to another even in the same $\rho$ class, which may be caused by the change of the accretion rate , magnetic field or winds. It will be useful to compare the long-timescale light curve of HLX-1 with those of GRS 1915+105 and IGR J17091-3624, which can be used to distinguish the disk-instability model with other models.

\subsection{The universal correlation between bolometric luminosity and outburst duration}
 The radiation-pressure instability model predicts that the outburst duration, $T_{\rm dur}$, is positively correlated with the bolometric luminosity, $L_{\rm bol}$, for a given viscosity parameter \citep[][]{czerny09}. Here, we explore the possible correlation of $L_{\rm bol}-T_{\rm dur}$ for BH masses covering many orders of magnitude (BH XRBs, HLX-1 and AGNs) that possibly triggered by radiation-pressure instability. We neglect the neutron star-MXB 1730-335 due to there may be some quantitative differences bewteen BHs and neutron stars. For stellar-mass BHs, we adopt two sources that show regular heartbeat oscillations (GRS 1915+105 and IGR J17091-3624), where heartbeat oscillations have been observed in many data sets for both sources \citep[e.g.][]{neilsen12}. In this work, we select two typical X-ray observations with the outburst periods more or less equal to the mean values for each source (ObsID:20402-01-34-01, 62 outbursts in total exposure of $\sim$ 3400 s for GRS 1915+105 and ObsID:96420-01-05-04, 76 outbursts in total exposure $\sim$ 2850 s for IGR J17091-3624), which will not affect our main conclusion. The HLX-1 is selected as the IMBH candidate as discussed in Section 2.1. For AGNs, we select a possible candidate of Seyfert galaxy IC 3599 and the young radio galaxies (see Introduction), where these sources are possibly triggered by radiation-pressure instability \citep[][]{czerny09,wu09}.  We select 37 young radio galaxies (23 CSS and 14 GPS) from \citet{wu09}. In total, our sample includes 2 XRBs, HLX-1 and 38 AGNs (37 young radio galaxies and IC 3599, see Table 1).  
        
To estimate the bolometric luminosities of two BH XRBs, we reduce the selected X-ray data for GRS 1915+105 (ObsID:20402-01-34-01) and IGR J17091-3624 (ObsID:96420-01-05-04). The data reduction for the \emph{RXTE}/PCA data was carried out with HEASoft version 6.8 following standard analysis procedure. We use the 3-35 keV energy range for \emph{RXTE}/PCA spectral analysis since the source count-rate statistics are poor outside this range (e.g., 35-60 keV). While fitting, we add 1\% systematic errors. We try to use as simple model as possible to fit the X-ray data, and we find the X-ray data can roughly fitted with a model consisting of a \texttt{PHABS}, a \texttt{diskbb}, a \texttt{broken powerlaw} and a \texttt{Gaussian}. In the fitting, we add a small Gaussian component at 6.4 keV and adopt a broken power-law model to improve the reduced $\chi^2$, due to the simple power-law model yields unacceptably large reduced $\chi^2$ ($>$1.5). The absorption column densities were fixed at $N_{\rm H}=6.0\times10^{22} \rm cm^{-2}$ and $0.9\times10^{22} \rm cm^{-2}$ for GRS 1915+105 and IGR J17091-3624 respectively \citep[e.g.,][]{pahari14}. We list fitting results in Table (2). With the model parameters, we estimate the 0.1--200 keV luminosities for each XRB and adopt the luminosity in this waveband as an approximation of bolometric luminosities, which are $9.6^{+3.2}_{-3.1}\times10^{38}\rm erg\ s^{-1}$ and $4.6^{+2.1}_{-1.8}\times10^{37}\rm erg\ s^{-1}$ for GRS 1915+105 and IGR J17091-3624 for the distances of $12.5\pm2.1$ kpc and $14\pm3$ respectively \citep[][]{muno99,rodr11}. HLX-1 is a highly variable source with 0.3--10 keV X-ray luminosities ranging from  $2.1\times10^{40}$ to $1.3\times10^{42}\rm erg\ s^{-1}$. Assuming the bolometric correction of $\sim 5$ for the 0.3-10 keV X-ray luminosity as in XRBs and AGNs \citep[][]{macc03}, the bolometric luminosity is $\sim1.1\times10^{41}$ at the low state to $6.5\times10^{42}\rm erg\ s^{-1}$ at high state. Here, we adopt $\log L_{\rm bol}=41.9^{+0.9}_{-0.9}$ as an approximation. For IC 3599, the mean bolometric luminosity is $\log L_{\rm bol}=43.5^{+1.1}_{-1.1}$, which is estimated from 0.3-10 keV X-ray luminosities \citep[][]{gru15} assuming a bolometric correction factor of 5 \citep[][]{macc03}, where the highest and lowest X-ray luminosities are $9.12\times 10^{43}$ and $4.7\times 10^{41}\rm erg\ s^{-1}$ respectively. For the young radio galaxies, the bolometric luminosities are selected from literatures \citep[][]{wu09,wu09b}, which are estimated either from the optical luminosities at given waveband or from the optical-emission-line luminosities using the empirical correlations between line luminosity and bolometric luminosity. We statistically take these luminosity values as the mean luminosity of each source.

We define the outburst duration as the timescale between the luminosity increases and decreases to $(L_{\rm max}+L_{\rm min})/2$ respectively, where $L_{\rm max}$ and $L_{\rm min}$ represent the maximum and minimum luminosities in each outburst. %We fit the light curves of GRS 1915+105, IGR J17091-3624, HLX-1 and IC 3599 using an exponential rise ($L_{\rm r}(t)=\rm a e^{\it t/t_{\rm r}}$) and an exponential decay function ($L_{d}(t)=\rm a e^{\it -t/t_{\rm d}}$).
The average outburst durations are $9.8\pm0.2$ s (1 $\sigma$), $4.5\pm0.2$ s, $54.9\pm20.4$ days and 2 years for 62 outbursts of GRS 1915+105, 76 outbursts for IGR J17091-3624, 6 outbursts of HLX-1 and 1 outburst of IC 3599 respectively. For the compact radio galaxies, both the kinematic ages estimated from the proper motions of radio lobe and the radiative ages derived from the synchrotron emission based on the radio spectra of radio lobes support that the compact radio sources are young ($\sim10^{2-5}$ years). In this work, we statistically regard the age of young radio galaxies as the outburst duration, where each outburst is associated with the ejection of radio jets in the active phase, and the ages of these young radio galaxies are mainly selected from literatures \citep[e.g.,][]{giro09,wu09}.    

In Figure 2, we present the relation between the duration of outburst and bolometric luminosity for these different-scale BHs (2 XRBs, HLX-1 and 38 AGNs), where these two quantities are positively correlated . The best fit is $\log L_{\rm bol}=0.70\pm0.02 \log T_{\rm dur} + 43.01\pm0.13$ (solid line), which is roughly consistent with that predicted by the radiation-pressure instability of accretion model (dashed line, $\log L_{\rm bol}=0.80 \log T_{\rm dur}+42.88$ for the case of $\alpha=0.02$, \citealt{czerny09}). The universal correlation between the duration and the bolometric luminosity in XRBs, HLX-1 and AGNs suggests that the radiation-pressure instability may exist in different-scale BHs, which establishes a universality of accretion physics. It should be noted that only one typical observation was selected for each XRB, and the oscillation period and duration are not exactly the same from one observations to another one. It was found that the period is $\sim 60\pm20$ s (1 $\sigma$) for different observations in GRS 1915+105 \citep[e.g.,][]{neilsen12}, and, therefore, the difference is not so large and it will not affect our main conclusion on the universal correlation.

It is still unclear, why the heartbeat oscillations were only evident in two BH XRBs (GRS 1915+105 and IGR J17091-3624). Recently, it was found that heartbeat oscillations may also exist in other BH XRBs in disk dominated soft state and intermediate states \citep[][]{sukova16}, even though their light curves are less regular compared to those found in heartbeat states of GRS 1915+105 and IGR J17091-3624. It suggests that the radiation-pressure instability may only clearly appear when the accretion rate is close to the Eddington rate. The magnetic field, wind and corona can stabilize the disk and suppress the oscillations, or at least make them less pronounced \citep[][]{czerny03,zheng11,janiuk15,sad16}, which will lead to a much higher critical value of the Eddington rate than that of the pure disk. Since the strength of those stabilizing mechanisms is difficult to estimate, the observational approach to the presence or absence of the radiation pressure instability has a clear advantage. We calculate the Eddington ratios for the sources in our sample with estimated BH masses and bolometric luminosities. We find that $L_{\rm bol}/L_{\rm Edd}$ range from $\sim10^{-2}$ to several with a distribution peak of $\sim 0.3$ (see Figure 3). For supermassive BHs, the radiation-pressure instability will lead to strong variations with amplitude of two orders of magnitude during the outbursts, where the Eddington ratio can be around $10^{-2}$ to several even if the average accretion rate may be just close to Eddington rate \citep[][]{czerny09}. For stellar-mass BH of IGR J17091-3624, the Eddington ratio is about one order of magnitude lower than that of GRS 1915+105. The faintness of IGR J17091-3624 may be caused not by the low accretion rate but by a very low or negative spin of BH \citep[e.g.,][]{rao12}, while GRS 1915+105 has a near maximally spinning BH \citep[][]{mcc06}. Further test in the radiation-pressure instability model under Kerr BH should be useful to test this issue. It should be noted that most of BH systems in our sample have quite high bolometric Eddington ratios (average value of $L_{\rm bol}/L_{\rm Edd}$ is around 0.3, $L_{\rm Edd}$ is Eddington luminosity).

\section{Summary}
   The HLX-1 is the best candidate for the IMBH, which underwent strong periodic outbursts in last several years after it was monitored by \textit{Swift}. The physical reason for the recurrent outbursts is still unclear. We model the light curve of HLX-1 with the time evolution of the accretion disk under the radiation-pressure instability, where this mechanism has been briefly discussed in \citet{lasota11} and \citet{sun16}. We also explored the possible universal properties of the different-scale BHs that may be regulated by radiation-pressure instability. Our main results include:\\
 1) The radiation-pressure instability in accretion disk with an IMBH of $\sim 10^5 \msun$ can roughly reproduce the recurrent outbursts of HLX-1 (e.g., period, duration and amplitude of the outbursts), which suggest that the outbursts of HLX-1 may be indeed driven by radiation-pressure instability (see Figure 1).\\
 2) We find a universal correlation between bolometric luminosity and outburst duration for BHs at different scales (XRBs, HLX-1 and AGNs), where this correlation is roughly consistent with that predicted by the model of radiation-pressure instability (see Figure 2). It  establish a new universality of accretion physics.

\section*{Acknowledgements}
We thanks S. Komossa (MPIfR) for suggesting the Seyfert galaxy of IC 3599, and Wenfei Yu (SHAO), Yan-Fei Jiang (Harvard-Smithsonian Center for Astrophysics), and Bing Zhang (UNLV) for very helpful discussions. Authors acknowledge support from the natural Science Foundation of China (NSFC, QW: 11573009, 11133005, XFC:11303010, WMG: 11573023 and ZY: 11403074), the Polish National Science Center (AJ and MG: DEC-2012/05/E/ST9/03914, BC: 2015/17/B/ST9/03436) and FP7/2007-2013/312789 (BC).

% -----------------------------------------------------------
\newpage
\begin{table*}
\centering
\centerline{Table 1: Relevant parameters of the BH sources.}
\footnotesize
\begin{tabular}{lccccccc}
\hline
\hline
Name & $z$  &$T_{\rm dur}$ &$\log L_{\rm bol}$  & $\log M_{\rm BH}$ & $\log L_{\rm bol}/L_{\rm Edd}$  & Type  & References \\
     &           & year         &  erg/s              & $\msun$          &              &            &  \\
~[1] &[2]  &[3]           &[4]  &[5]  &[6]    &    [7]  & [8] \\
\hline
\hline
   GRS 1915+105& ...& $3.1^{+0.1}_{-0.1}\times10^{-7}$ & $39.0^{+0.1}_{-0.2}$ & $1.0^{+0.02}_{-0.02}$ &-0.2    & XRB   & 1,2    \\
   IGR J17091-3624 & ...& $1.4^{+0.1}_{-0.1}\times10^{-7}$ & $37.7^{+0.2}_{-0.2}$ & $0.9^{+0.2}_{-0.3}$ & -1.4  & XRB & 1,3\\
   HLX-1      &  0.022  & $0.15^{+0.05}_{-0.04}$ & $41.9^{+0.9}_{-0.9}$ & 5.1 & -0.5 & HLX & 1,1  \\
   IC 3599    &  0.022  &   2.0      & $43.5^{+1.1}_{-1.1}$ & 6.3 & -1.0 & Seyfert& 1,4   \\
   0108+388   &  0.669  &   400      & 44.2  & 7.9  &   -1.8   & GPS   & 5,5\\
   0710+439   &  0.518  &   930      & 45.8  & 8.4  &   -0.8   & GPS   & 5,5\\
   1031+567   &  0.450  &   1840     & 45.0  & 8.1  &   -1.2   & GPS   & 5,5\\
   1358+624   &  0.431  &   2400     & 45.1  & 8.2  &   -1.3   & GPS   & 5,5\\
   1404+286   &  0.077  &   220      & 45.2  & 8.7  &   -1.7   & GPS   & 5,5\\
   1934-638   &  0.183  &   1600     & 45.6  & 8.5  &   -1.0   & GPS   & 5,5\\
   2021+614   &  0.227  &   370      & 45.2  & 8.9  &   -1.9   & GPS   & 5,5\\
   2352+495   &  0.238  &   3000     & 44.7  & 8.4  &   -1.8   & GPS   & 5,5\\
   J1111+1955 &  0.299  &   1620     & 45.1  & 8.5  &   -1.5   & GPS   & 5,5\\
   0116+31    &  0.060  &   500      & 44.9  &  ... &     ...  & GPS   & 5,5\\
   1718-649   &  0.014  &   90       & 44.1  & 8.4  &   -2.4   & GPS   & 5,5\\
   0316+413   &  0.018  &   240      & 44.6  & 8.5  &   -2.0   & GPS   & 5,5\\
   1413+135   &  0.247  &   130      & 44.1  & 8.0  &   -1.9   & GPS   & 5,5\\
   0035+227   &  0.096  &   450      & 44.9  & ...  &     ...  & GPS   & 5,5\\
   0221+276   &  0.310  &   51000    & 46.3  & 8.1  &    0.0   & CSS   & 5,5\\
   0404+769   &  0.599  &   1100     & 45.3  & 8.3  &   -1.1   & CSS   & 5,5\\
   0740+380   &  1.067  &   113000   & 46.8  & 8.9  &   -0.3   & CSS   & 5,5\\
   1019+222   &  1.617  &   9300     & 46.7  & 7.8  &    0.6   & CSS   & 5,5\\
   1203+645   &  0.371  &   35400    & 46.1  & 7.8  &    0.2   & CSS   & 5,5\\
   1250+568   &  0.320  &   204000   & 45.9  & 7.6  &   -0.1   & CSS   & 5,5\\
   1416+067   &  1.437  &   70000    & 47.6  & 10.4 &   -0.8   & CSS   & 5,5\\
   1443+77    &  0.267  &   110000   & 46.4  &  8.4 &   -0.2   & CSS   & 5,5\\
   1447+77    &  1.132  &   151000   & 47.7  &  8.9 &    0.6   & CSS   & 5,5\\
   2252+12    &  0.543  &   348000   & 46.6  &  8.5 &   -0.0   & CSS   & 5,5\\
   2342+821   &  0.735  &   13000    & 45.8  &  7.5 &    0.2   & CSS   & 5,5\\
   0127+233   &  1.459  &   200000   & 46.3  &  9.2 &   -1.0   & CSS   & 5,5\\
   0134+329   &  0.367  & $>$10000   & 46.6  &  9.2 &   -0.3   & CSS   & 5,5\\
   0518+165   &  0.759  &   50000    & 46.4  &  8.7 &   -0.4   & CSS   & 5,5\\
   0538+498   &  0.545  & $>$39000   & 45.9  &  8.7 &   -0.9   & CSS   & 5,5\\
   0758+143   &  1.195  &   50000    & 45.9  &  7.8 &   -0.1   & CSS   & 5,5\\
   1328+307   &  0.849  &   462000   & 46.6  &  8.5 &   -0.1   & CSS   & 5,5\\
   1328+254   &  1.055  &  55000     & 46.9  &  9.6 &   -0.9   & CSS   & 5,5\\
   1458+718   &  0.905  & $>$13000   & 46.9  &  9.1 &   -0.3   & CSS   & 5,5\\
   2249+185   &  1.757  & $>$3400    & 46.1  &  8.6 &   -0.6   & CSS   & 5,5\\
   0345+337   &  0.243  &   35800    & 45.6  &  7.1 &    0.4   & CSS   & 5,5\\
   1517+204   &  1.574  &   150000   & 45.9  &  ... &    ...   & CSS   & 5,5\\
   0809+404   &  0.551  &   100000   & 46.0  &  8.8 &   -0.9   & CSS   & 5,5\\

\hline
\hline
\end{tabular}
\footnotesize
\begin{minipage}{170mm}
%Col. 2: Redshift, the distance of $12.5\pm2.1$ kpc and $14\pm3$ kpc are adopted for GRS 1915+105 and IGR J17091-3624 respectively;
References for ($T_{\rm dur}$, $L_{\rm bol}$) and $M_{\rm BH}$: 1) this work; 2) Steeghs et al. 2013; 3) Pahari et al. 2014; Rebusco et al. 2012; 4)Grupe et al. 2015; 5) Wu et al. 2009a and references therein.
\end{minipage}
\label{para}
\end{table*}

\begin{table*}
\centering
\centerline{Table 2: The best-fit parameters for GRS 1915+105 and IGR J17901-3624.}
\begin{tabular}{lccc}
\hline
\hline
Model & Parameter &  GRS 1915+105   & IGR J17901-3624 \\
\hline
\hline
phabs  & $N_{\rm H}$($\times10^{22}\rm cm^{-2}$)     &   6.0 (fix)         & 0.9 (fix)                  \\
diskbb & $\kappa T_{\rm in}$ (keV)                   & $1.40\pm0.04$       & $1.18\pm0.02$                \\
diskbb & $N_{\rm diskbb}$                            & $314.24\pm56.02$    & $26.64\pm3.08$               \\
Gaussian Line & LineE (keV)                          &  6.4(fix)           & 6.4 (fix)                     \\
Gaussian Line & $\sigma$ (keV)                       &  $1.51\pm0.25$      & $0.44\pm0.28$                  \\
Gaussian Line & $N_{\rm line}$                       & $(2.88\pm1.44)\times10^{-2}$ & $(5.61\pm2.47)\times10^{-4}$       \\
bknpower      & $\Gamma_{1}$                         &  $2.55\pm0.06$      & $2.51\pm0.09$         \\
bknpower      & $E_{K}$ (keV)                        &  $13.01\pm0.51$     & $15.32\pm3.11$        \\
bknpower      & $\Gamma_{2}$                         &  $2.95\pm0.02$      & $2.07\pm0.33$        \\
bknpower      & $N_{\rm pl}$                         &  $16.62\pm2.50$     & $0.48\pm0.10$         \\
$\chi^2$/dof  &                                      & (54.39/57)0.95      & (40.71/53)0.77     \\

\hline
\hline
\end{tabular}
\footnotesize
\begin{minipage}{170mm}
We try to estimate the 0.1-200 keV luminosity as the bolometric luminosity based on SED modeling for these two XRBs. For this purpose, we adopt the model as simple as possible, which is $phabs*(diskbb+gauss+bknpower)$.
\end{minipage}
%\label{para}
\end{table*}

\begin{figure}
\epsscale{1.0} \plotone{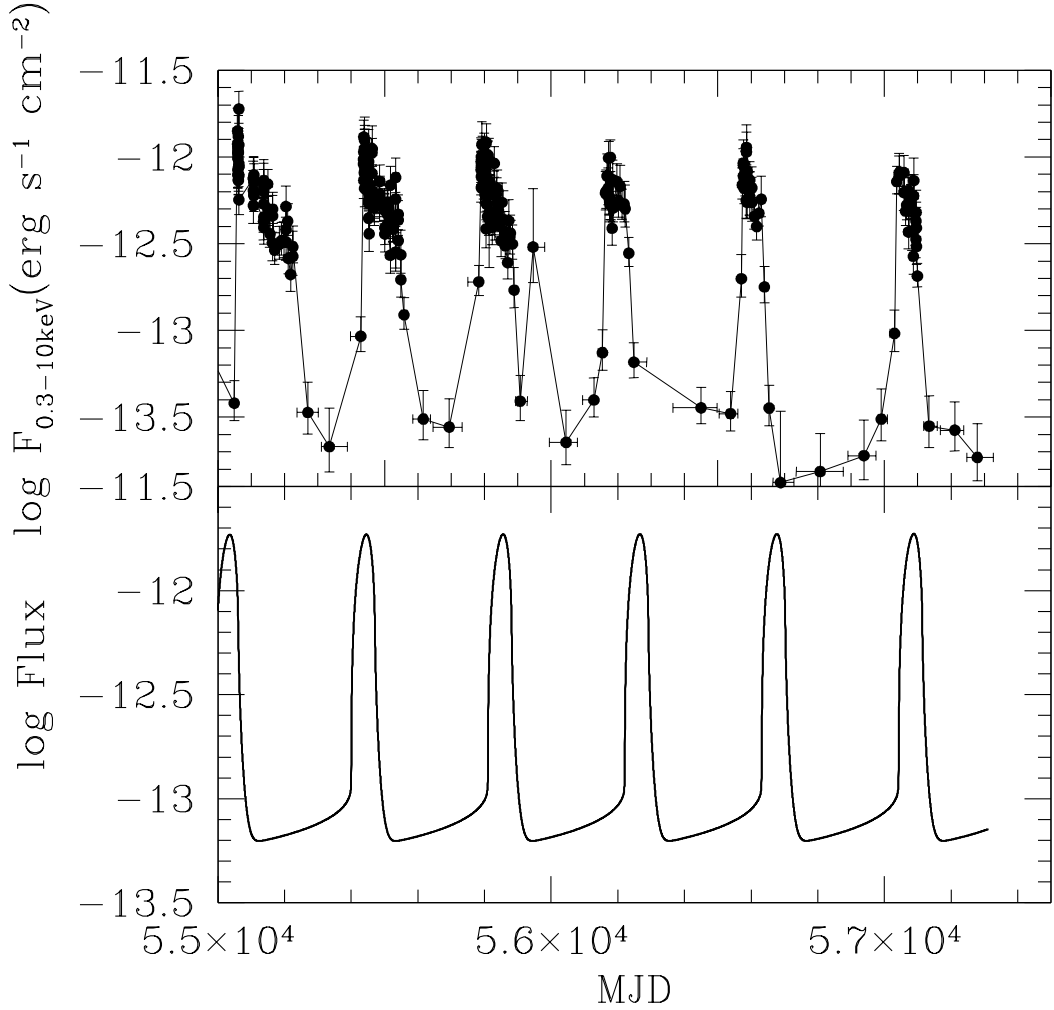} \caption{Observational light curve of HLX-1 (top panel) and modeled light curve of accretion disk (bottom). The model flux at the same energy band is derived from bolometric flux by assuming Flux=$F_{\rm bol}/5$, where $F_{\rm bol}=L_{\rm bol}/4\pi D_{\rm L}^2$ ($D_{\rm L}$ is the luminosity distance of HLX-1) and 5 is the bolometric correction factor \citep[][]{macc03}.}
\end{figure}

\begin{figure}
\epsscale{1.0} \plotone{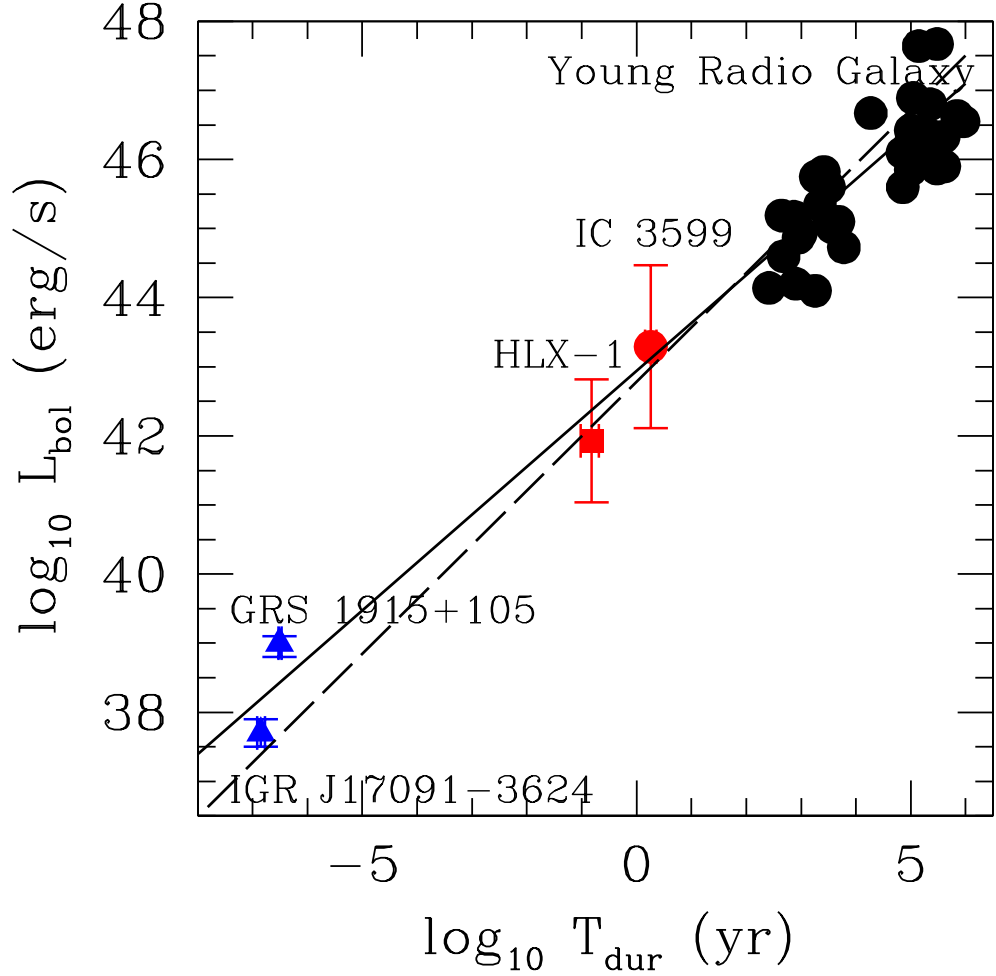} \caption{Correlation between the bolometric luminosity and the outburst duration for different-scale BHs. The solid line represents the best fit and the dashed line shows the prediction of the disk model under radiation-pressure instability ($\alpha=0.02$, \citealt{czerny09}). }
\end{figure}

\begin{figure}
\epsscale{1.0} \plotone{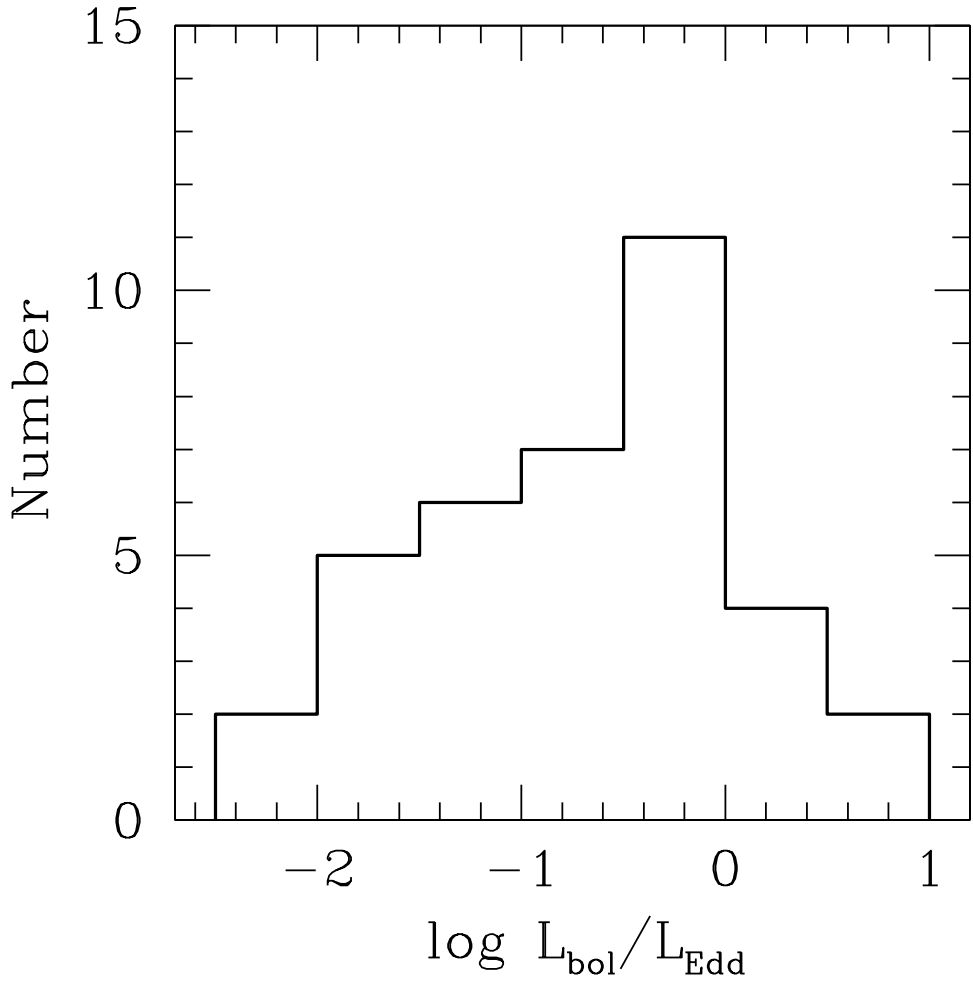} \caption{The distribution of Eddington ratios for BHs in our sample. }
\end{figure}

\end{document}